\documentclass[10pt,journal,compsoc]{IEEEtran}

\usepackage{tikz} 
\usetikzlibrary{positioning}
\usetikzlibrary{arrows,decorations.pathmorphing,backgrounds,fit,petri}
\usepackage[autostyle]{csquotes}
\usepackage{placeins}
\usepackage{graphicx}
\usepackage{subcaption}
\usepackage[justification=centering]{caption}
\usepackage{amsthm}

\usepackage{amsmath}
\usepackage{enumitem}
\usepackage{multirow}
\usepackage{subcaption}
\usepackage{algorithm}
\usepackage{algpseudocode}
\algrenewcommand\algorithmicrequire{\textbf{Input:}}
\algrenewcommand\algorithmicensure{\textbf{Output:}}

\SetLabelAlign{LeftAlignWithIndent}{\hspace*{1.0ex}\makebox[1.25em][l]{#1}}

\ifCLASSOPTIONcompsoc
  \usepackage[nocompress]{cite}
\else
  \usepackage{cite}
\fi

\ifCLASSINFOpdf
\else
\fi
\begin{document}
%
\title{Scheduling and Trade-off Analysis for Multi-Source
	Multi-Processor Systems with Divisible Loads}
%
%
%
%

\author{Yang~Cao,~\IEEEmembership{Stony Brook University,}
        Fei~Wu,~\IEEEmembership{Stony Brook University,}
        and~Thomas~Robertazzi,~\IEEEmembership{Fellow,~IEEE}

\IEEEcompsocitemizethanks{\IEEEcompsocthanksitem Y. Cao, F. Wu and T. Robertazzi are with the Department
	of Electrical and Computer Engineering, Stony Brook Univesity, Stony Brook,
	NY, 11794.\protect\\
	E-mail: yang.cao@stonybrook.edu}}

\IEEEtitleabstractindextext{%
\begin{abstract}
 The main goal of parallel processing is to provide users with performance that is much better than that of single processor systems. The execution of jobs is scheduled, which requires certain resources in order to meet certain criteria. Divisible load is a special but widely used type of data which can be divided into arbitrary sizes and independently processed in parallel. It can be used in applications which are processing a great amount of similar data units.  \par 
In this paper, a parallel system which has multiple sources and multiple processors is studied and the finish time/ makespan optimization problem is solved. In the system, load is distributed sequentially from the sources to the children processors. Both the scenarios of processing nodes with or without front-ends are considered. Numerical tests and simulations are done for both. A monetary cost model is proposed for estimating the overall computing cost of the system. This paper also discusses the trade-off relationship between monetary cost and minimal finish time. Detailed suggestions are given for the user who has a time budget, a cost budget or both.
\end{abstract}

}

\maketitle

\IEEEdisplaynontitleabstractindextext

%
\IEEEpeerreviewmaketitle

\IEEEraisesectionheading{\section{Introduction}\label{sec:introduction}}

\IEEEPARstart{P}{arallel} processing (or parallel computing) is a field in electrical engineering
and computer science related to the application of many computers running
in parallel to solve computationally intensive problems. The main goal of
parallel processing is to provide users with performance which no single
computer may deliver. Scheduling is an important task allowing parallel
systems to perform efficiently and reliably. In general, scheduling can be
considered as managing the execution of jobs which required certain resources
in such way that certain optimality and/or feasibility criteria are met.
Such optimality metrics can be minimal finish time, lowest monetary
cost and so on.\par
Divisible load is a special but widely used type of data which can be
divided into arbitrary sizes and independently processed in parallel. The
divisible loads may be commonly encountered in applications which are
processing a great amount of similar data units. During the past decades, Divisible
Load Theory (DLT) has been proved as a powerful tool for scheduling
in parallel systems.\par
This paper mainly studies a parallel system which has multiple
sources and multiple processors. Sequential load distribution is used for the workload distributing procedure. Compared with regular single source load
distribution systems, multiple sources have to be scheduled to communicate
with the processing nodes in a specific sequence which solves the finish
time optimization problem. When processing
nodes have front-end processors, the nodes can compute and communicate at the same
time. So both the scenarios of the processing nodes with or
without front-ends are considered. Numerical tests and simulations show that the multi-source
multi-processor system has significant improvement compared with
single-source systems by reducing the system minimal finish time.\par
In this paper, a monetary cost model is also proposed for estimating the
overall computing power used by the system. The trade-off relationship
between monetary cost and minimal finish time is discussed for different situations. Detailed suggestions are given for the user who has a time budget, a cost budget or both.

\subsection{Background}
In the past decade, parallel and distributed systems have become a very
general application. To process large-scale, data-intensive loads, multiple
processors are required to work in parallel. The most important task for a
scheduling problem is to assign different amounts of data to these parallel
computers and make them finish each partition in an acceptable temporal
range. Parallel systems are often used in the areas that have heavy computation
requirements.\par
In order to study the processing of load for parallel and distributed computing, Divisible
Load Theory (DLT) was created [1] [5] [18] [19] [22]. It assumes that communication
and computation loads can be partitioned arbitrarily among numerous processors
and being processed in parallel. In 1988 [1], Cheng's paper first
gave an intuitive proof for the Divisible Load Theory’s optimality principle.
Five years later, a formal proof was given and an extensive search to validate
the result was run on an IBM mainframe. Since then, DLT was developed
and studied with multiple network topologies. Topologies include bus
networks [2], star networks, tree networks [3], meshes [4], grids [14] [15], etc. Nowadays,
DLT has also been developed for more different environments. These include
cloud networks [9] and sensor networks [7] [8]. It has become a
powerful tool for modeling data-intensive computational problems.\par

\subsection{Applications}
Potential applications of Divisible Load Theory can be widely found in the fields of
image processing, video processing, sensor networks, cloud networks, etc. The following section gives more details of these applications.\par 

\subsubsection{Image Processing}
Image feature extraction is a highly used function in computer vision systems. There are mainly two phases of computation for image feature extraction. In the first phase, the image will be segmented into many pieces and be processed independently and locally by different processors. During this procedure, the local features of the image will be extracted. In the second phase of computation, local features from different processors are exchanged and then processed to extract the desired features. The first phase of image feature extraction can be considered to use DLT since the load can be arbitrarily divisible since there is no precedence relations [5].\par
\subsubsection{Video Processing}
Another application for DLT is video processing. With the rapid growth of digital TV and interactive media over broadcast networks, the need for high performance computing for broadcasting is much more important than earlier. Parallel processing is one of the  best ways to meet the need for a considerable amount of data processing. The authors of [6]  first applied the DLT paradigm to the video encoding process and designed a parallel video encoder which was shown to achieve a good performance. With the help of DLT, the precise modeling and minimization of the execution time of each phase of the video encoding process becomes an easy task.\par
\subsubsection{Sensor Network}
Wireless sensor networks (WSN)  are spatially distributed autonomous sensors to monitor physical or environmental conditions, such as temperature, sound, pressure, etc, and then cooperatively pass their data through the network to a main location. 

The problem of load distribution in a large scale sensor network was defined as an optimization problem to minimize the overall finish time of the whole system [7] [8]. By finishing sensing tasks faster, a system could get the returned results more quickly and also save more energy and monetary cost. Since the data collected by multiple sensors may have no precedence relations, it can be considered as a divisible load. In this case, DLT can be applied to sensor network applications to improve their performance.\par

\subsubsection{Cloud Network}

Cloud computing is an on demand service in which shared resources, information, software and other devices are provided according to the clients requirement at specific time [9].
 Cloud network provides continuity for large-scale service-oriented applications [10] [11]. For more details on cloud computing, refer to [12] [13] [16] [17].The users may require the whole cloud network to process a job which is very data-intensive. To process the job, efficient load balancing techniques are needed, which involves reassigning the total load to the individual nodes of the collective systems to make resource utilization effective and to improve the response time of the job. The Divisible Load Theory paradigm is a very powerful tool for solving the load balancing problem, as long as the load is arbitrarily divisible.\par

\subsection{Motivations and Contributions}

In most of the previous work of Divisible Load Theory, it is often assumed that there is only one source to store the original data. This source node may deliver the data fractions to each processing node in a sequential manner, which leads to a result that many processing nodes are idle when they do not have any data to process. This results in a waste of computing resources and lower efficiency. Nowadays, with the rapid development of network and cloud computing, it is very practical to store the original data in different databanks and later send it to different processors for further computation [17].

It is also very practical to adapt the multi-source topology to traditional networks. In this paper, the topology of two-level tree networks fed data by a data originator node is considered. The original data is stored in the data originator on the first layer, which is only one source node. 

 In the second layer of the tree topology, there are a few source nodes, which receive load from the first layer and further transmit them to the third layer. Finally, the third layer contains many processing nodes that do the computing tasks in parallel.\par
This paper is mainly focused on the last two layers of this two-level tree network. Compared with the previous work on multi-source systems, the study is separated into two cases, which is that the processors have front-end processors, or the processors are not equipped with front-end processors. Closed-form solutions are found for both of the scenarios to achieve the overall minimal finish time for the system. Moreover, a monetary cost model is developed to estimate the cost charged by using the processors' computing power. The trade-off between monetary cost and system finish time/ makespan will be given. More suggestions are discussed for the users who have monetary budgets to use the system, or have requirements to finish the task in a certain time range.
\par

\subsection{Organization}
The rest of this proposal is organized as follows. Section 2 first briefly introduces the basics of a classic scheduling problem using Divisible Load Theory. Then in section 3, a multi-source multi-processor network topology is studied, which is divided by two cases: the processing nodes are equipped with front-end processors or the processing nodes are not equipped with front-end processors. Section 4 contains the numerical tests and simulations for this parallel distribution system. The system speedup and performance analysis using Amdahl's Law will be shown in section 5. In section 6, we investigate calculating the overall monetary cost of the distribution system. More detailed discussion of the trade-off between the minimal finish time and monetary cost is covered here. The conclusion and future works appears in section 7 and 8.\par
\par
The following notation is used in this proposal:
\begin{enumerate}
	\item[$\beta_{i,j}$] The fraction of divisible load that is assigned from source $S_i$ to processor $P_j$.
	\item[$G_{i}$] The inverse communicating speed of source $S_i$.
	\item[$R_{i}$]  The release time of source $S_i$.
	\item[$A_{j}$] The inverse computation speed of processor $P_j$.
	\item[$C_{j}$] The cost for processor $P_j$ to work for one unit of time.
	\item[$T_{f}$] The finish time of processing the entire job.
	\item[$J$] The total job (amount of data) that needs to be distributed and processed.
	\item[$Cost_{total}$] The overall monetary cost for the entire system.
	\item[$Budget_{cost}$] The user's budget for the monetary cost.
	\item[$Budget_{time}$] The user's budget for the finish time.
\end{enumerate}

\section{Basics for Divisible Load Theory (DLT)}
This section mainly studies the basics for Divisible Load Theory. A fundamental model is given and a closed-form solution for the overall minimal finish time is presented. 
\subsection{Definition}
Divisible Load Theory (DLT) is a methodology involving the linear and continuous modeling of partitionable computation and communication loads for parallel processing. There is a fundamental assumption for most of the divisible load studies. In order to achieve the minimal finish time, all of the processors should finish processing the fractions of load that are assigned to themselves at the same time. If not, the unfinished data can always be sent to the processors which already finished processing. A formal proof of this assumption was given in [2].

\subsection{Problem Formulation}
\begin{figure}[h!]
	\centering
	\includegraphics[width=0.52\textwidth]{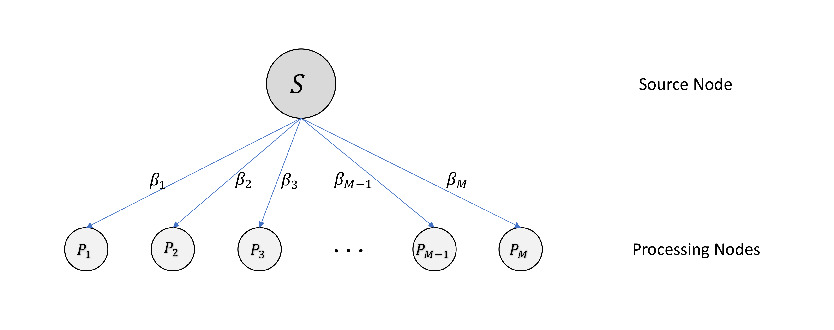}
	\caption{A Load Distribution System with One Source and M Processors}
	\label{fig:ts}
\end{figure}	
A basic model using DLT is shown in Figure 1. The topology is a single-level tree network. The top layer is a source node, which stores all the data after getting a work task. It transmits different amount of load partitions to the second layer of the tree, where there are M processing nodes. These processors can do parallel computing once each of them receive their fraction of load  $\beta_j$. Each processor has a separate link to connect the source node for communication. Based on the assumption that the source can only communicate with one processor at a time, the arrangement of the communication between source and processors is as follows: \par
First of all, the source node $S$ does sequential communication, which means that it communicates with the arranged  order of processors {$P_1$, $P_2$, $P_3$, ...,  $P_{M-1}$, $P_M$}. Secondly, to achieve shorter finish time of processing the whole task, the processing nodes are sorted by the descending order of their computing speed [5], which is $A_1 \leq A_2\leq A_3\leq...\leq A_{M-1}\leq A_M$ (please note that $A_j$ is the inverse computing speed of $P_j$). This makes the processors with faster computing speed start processing earlier than the ones with slower speed. As another result, the time that all the processors finishes processing, which is called  finish time $T_f$ can be shorter.\par 
The main problem to solve is to find a load distribution plan that minimizes the overall system finish time $T_f$, which includes all the processing nodes as well as the source node. \par 
\begin{figure}[h!]
	\centering
	\includegraphics[width=9.5cm]{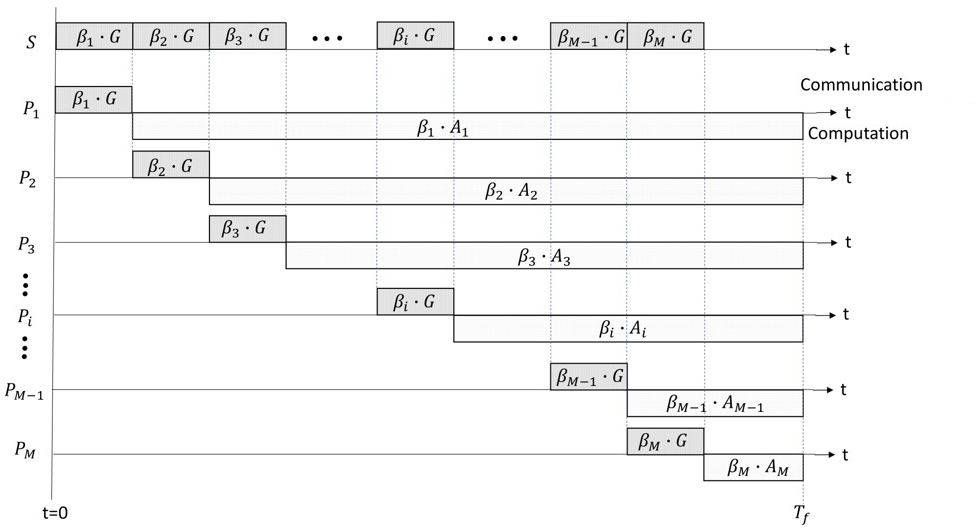} 
	\caption{Timing Diagram for a Single-level Single-source Tree Network (without Front-End Processors)}
	\label{fig:ts}
\end{figure}	
Figure 2 is the timing diagram for this basic load distribution system.
Here, $G$ denotes the inverse communication speed of source $S$ and $A_i$ is the inverse computing speed of processor $P_i$. The load fraction that source $S$ sends to processor $P_i$ is represented by $\beta_i$. The source node is not involved in the computing procedure. Each of the processors starts computing their fraction of load after finishing receiving it . They stop processing at the same time instant $T_f$ to achieve the minimal finish time of the whole system.\par
There are several assumptions for the processors: Firstly, a processor can only compute after it has finished the communication unless it is equipped with a front-end processor.Secondly,The source can only communicate with one worker processor at a time.Lastly,There is no communication between the worker processors.\par

The timing diagram indicates that source $S$ has continuous communication with the processors. Once it finishes sending load to processor $P_i$, it continues sending load to processor $P_{i+1}$. Since all of the processors finish processing at the same time, the following equations are written to represent the finish time for each processor. For processor $P_i$, its finish time equals the communication time that source communicated with $P_1$ to $P_{i-1}$, plus the computing time of $P_i$:
\begin{eqnarray}
T_{f} = \sum_{k = 1}^{i} \beta_{k}G + \beta_{i}A_{i}, \hspace{0.2cm} i = 1, 2,... , M
\end{eqnarray}
Also, based on the definition that $\beta_{i}$ is the fraction of load that source $S$ sends to processor node $P_{i}$, the following equation can be written to normalize the total amount of load which is processed by this system:\par
\begin{eqnarray}
\sum_{i = 1}^{M}\beta_{i} = 1
\end{eqnarray}
Since there are $M+1$ unknowns and $M+1$ linear equations, load partitions $\beta_{1}, \beta_{2},... ,\beta_{M}$ can be uniquely solved as well as the system finish time $T_{f}$.\par

\section[Scheduling for Multi-Source System]{Scheduling for Multi-Source Multi-Processor System}
In this section, a multi-source, multi-processor load distribution system is studied.  The study will be divided into two scenarios: the processing nodes with or without front-end professors. Here, front-end processor refers to a small-sized sub-processor which has the job of data collection and communication between source node and processing node. If a processing node is equipped with a front-end processor, it can start computing the data once it starts receiving it with the front-end processor.\par 
There is an assumption that it always take a much longer time to compute the data rather than transfer it. In this case, if the node continuous receiving data, it can achieve continuous processing, which is assumed to be more efficient and energy-wise.\par
There is also an assumption that the load which need to be sent by each source to the children processors has already been received from the job allocator by the time when each source starts distributing load.\par 
Meanwhile, since there are multiple sources ($S_1\sim S_N$) that are distributing load fractions in parallel, they are sorted in order to achieve shorter finish time. The system would always start using the sources which have faster communication speeds so that the processors could get the load fractions earlier. In this section, they are sorted in the descending order of their communication speed, which is  $G_1 \leq G_2\leq G_3\leq...\leq G_{N-1}\leq G_N$ (please be noted that $G_i$ is the inverse communication speed of $S_i$). In this paper, the link speeds is determined by the communication speeds of the sources.

\subsection{Scheduling with Front-end Processor}
\subsubsection{Network Topology}
\begin{figure}[h!]
	\centering
	\includegraphics[width=9cm]{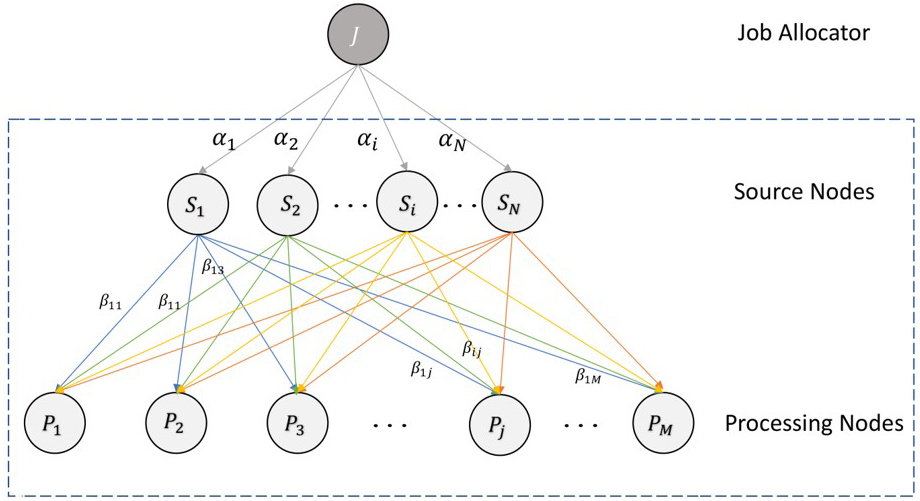} 
	\caption{Network Topology for a Multi-source Multi-processor Network}
	\label{fig:ts}
\end{figure}	
Figure 3 is the network topology for a multi-source multi-processor network. It is a two-level tree topology. Compared with the single source single-level tree network discussed with Figure 1, this network has one more layer, which is a job allocator/originator $J$ that stores all of the data that is needed for computing. This job allocator distribute fractions of load to the second layer, where there are N source nodes. Then, the source nodes further distribute the load into smaller fractions and allocate them to M processing nodes on the third layer.\par
 For each source node $S_i$, the amount of load it obtains from job allocator $J$ is denoted by $\alpha _i$ , which equals to the total load that $S_i$ sends to all the processing nodes. So the following equations can be written: $\alpha_i=\sum_{j = 1}^{M}\beta_{i,j}$. By solving the minimization problem of system finish time, all of the values for $\beta_{i,j}$, where $i=1, 2, 3, ..., N; j=1, 2, 3,. .., M$ can be found. So in order to simplify the problem, this paper mainly focuses on the two lower layers of this network.\par 
\subsubsection{Problem Formulation}

\begin{figure}[h!]
	\centering
	\includegraphics[width=9.2cm]{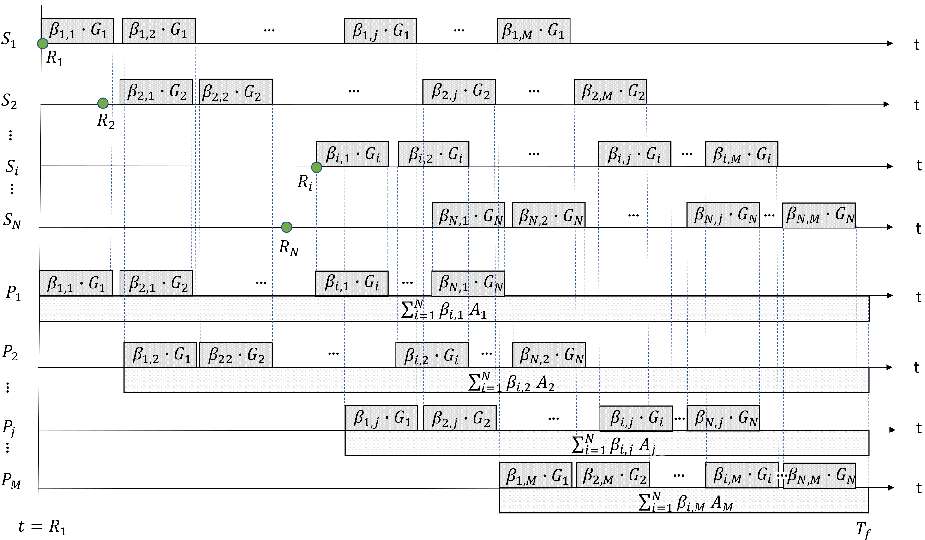} 
	\caption{Timing Diagram for a Multi-source Multi-processor Network (with Front-End Processors)}
	\label{fig:ts}
\end{figure}

The timing diagram for multi-source multi-processor distribution system is shown in Figure 4. Here,   $\beta_{i,j}$ denotes the fraction of load that source $G_{i}$ sends to processor $P_{j}$. For source $S_{i}$ it can start sending load to the $P_{1}$ right after the time reaches its release time $R_{i}$, or after the previous source $S_{i-1}$ finishes sending load to $P_{1}$, whichever is later.\par 
The order that each source distributes load fractions to processors is the same as the order that the processors are sorted, which means that processors with faster computing speed receive load earlier than the ones with slower computing speed. For the processors, the order that they receive load fractions from different sources matches the order that the sources are sorted.\par

Inspired by the previous work [17], the following part will discuss the constraints for this problem.\par 
 \textbf {A. Constraints Introduced by Release Time}\par 
	First, a new parameter is introduced for this system, which is called the release time of sources. It is denoted by $R_i$ and shows when source $S_i$ first become available for usage.\par 
	\begin{figure}[h!]
		\centering
		\includegraphics[width=9cm]{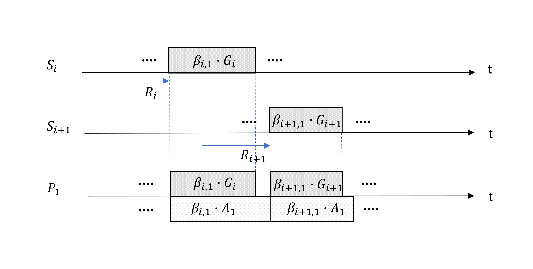}
		\caption{Timing Diagram for the First Processor Getting Load Fractions from Adjacent Sources}
		\label{fig:ts}
	\end{figure}

	Figure 5 shows the case that processor $P_{1}$ is getting load fractions from adjacent sources. The release times $R_{i}$ or $R_{i+1}$ may appear at any time point before the first fraction starting sending out from source $S_i$ or $S_{i+1}$.\par 
	In order to achieve continuous computing in $P_{1}$, the start time of sending load fraction {$\beta_{i+1,1}G_{i}$} should be exactly the same or earlier than the end time when $P_{1}$ finishes processing the previous load fraction. Also, the start time that source $S_{i+1}$ sends the first load fraction must be later or equal to its release time $R_{i+1}$.\par 
	 From the discussion above, the following criteria for release time can be proved:
	\begin{eqnarray}
	R_{i+1}\leq\ R_{i}+ \beta_{i,1}A_{1}, \hspace{0.4cm}    i = 1, 2,... , N-1
	\end{eqnarray}
	
\textbf {B. Constraints Introduced by Continuous Processing}\par
	The timing diagram is shown as Figure 4. It indicates that there might be some gaps between adjacent load fractions. In order to study them, they can be divided into two categories: gaps on sources, and gaps on processors.\par
	For gaps on sources, for example, load fraction {$\beta_{i,j}G_{i}$} and {$\beta_{i,j+1}G_{i}$} are the two fractions that source $S_{i}$, sends to processor $P_{j}$ and $P_{j+1}$ in a sequence. The gap may appear when the distribution of {$\beta_{i,j}G_{i}$}  is already finished while $P_{j+1}$ is still getting load fraction {$\beta_{i-1,j+1}G_{i-1}$} from source $S_{i-1}$.\par
	For gaps on processors, for example, load fraction {$\beta_{i,j}G_{i}$} and {$\beta_{i+1,j}G_{i+1}$} are the two fractions that processor $P_{j}$, gets from source $S_{i}$ and $S_{i+1}$ in a sequence. As discussed above, in order to be energy-wise, continuous processing for all the processors is required. The following constraints can be written based on Figure 6:\par
	\begin{eqnarray}
	\begin{aligned}
	\beta_{i,j}A_{j}+\beta_{i+1j,}G_{i+1}\leq\beta_{i,j}G_{i}+\beta_{i,j+1}A_{j+1}\\
	i = 1, 2,... , N-1, \hspace{0.4cm} j=1, 2, ..., M-1
	\end{aligned}
	\end{eqnarray}
	\begin{figure}[h!]
		\centering
		\includegraphics[width=9cm]{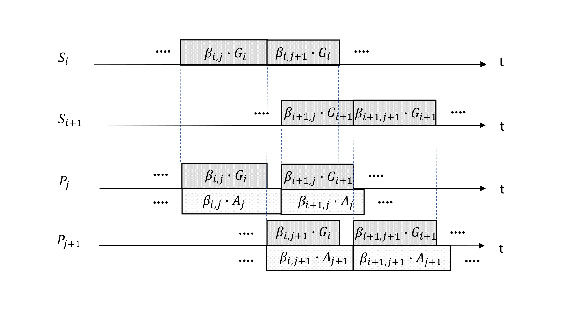}
		\caption{Timing Diagram for Adjacent Load Processing Tasks}
		\label{fig:ts}
	\end{figure}
	
	\textbf{C. Constraints Introduced by Finish Time}\par
	As the system timing diagram shows, the finish time for each processor equals the summation of two parts. The first part is the waiting time for processor $P_{j}$ to get the first load fraction, which is $R_{1}$ plus the time that $S_{1}$ distributes load fractions to processors $P_{1}$, $P_{2}$,..., $P_{j-1}$. The second part is the total processing time for the node to finish all the tasks. Since there might have many gaps during communication, the following  criteria can be written as an inequality:
	\begin{eqnarray}
	\begin{aligned}
	T_f\geq R_{1}+\sum_{k= 1}^{j-1}\beta_{1,k}G_{1}+\sum_{k= 1}^{N}\beta_{k,j}A_{j},\\
	j=1, 2, 3, ..., M
	\end{aligned}
	\end{eqnarray}
	
	\textbf{D. Constraints Introduced by Normalization}\par
	In order to normalize all the load fractions $\beta_{i,j}$, a parameter called total job $J$ is used:
	\begin{eqnarray}
	J=\sum_{i=1}^{N}\sum_{j=1}^{M}\beta_{i,j} 
	\end{eqnarray}

As a conclusion, an optimization problem is defined as the following:\par 
Given the number of sources (N), number of processors (M), each source sends load fractions to all the processors in a sequence, all the sources work in parallel, find the load fractions assigned to each processor from each source such that the total system finish time is minimized.\par 
Minimize $T_f$ such that:\par 
\begin{center}
	$R_{i+1}-R_{i}\leq\beta_{i,1}A_{1}, \hspace{0.4cm}    i = 1, 2, 3, ..., N-1$\\
	$\beta_{i,j}A_{j}+\beta_{i+1j,}G_{i+1}\leq\beta_{i,j}G_{i}+\beta_{i,j+1}A_{j+1}$, 
	\\$i = 1, 2,... , N-1,$ \hspace{0.4cm} $j=1, 2, 3, ..., M-1$\\
	$J=\sum_{i=1}^{N}\sum_{j=1}^{M}\beta_{i,j}$\\
	$T_f\geq R_{1}+\sum_{k= 1}^{j}\beta_{1,k}G_{1}+\sum_{k= 1}^{N}\beta_{k,j}A_{j}$,\hspace{0.4cm}$j=1, 2, 3,..., M$
\end{center}
In this problem, the variables are the system finish time $T_f$ and the load fractions $\beta_{i,j},\hspace{0.2cm} i=1, 2, 3, ..., N$ and $j=1, 2, 3, ..., M$. So it is a linear programming problem which has N * M + 1 variables.The solution of this problem is a point in a N * M + 1 dimensional space.\par

\subsection{Scheduling without Front-end Processor}
\subsubsection{Network Topology}
This section mainly discusses the case in which all the processing nodes are not equipped with front-end processors. In this case, a node can only start processing the data once all of it's data has been received. The network topology remains the same as Figure 3, however the timing diagram is changed as Figure 7 shows.\par 

\begin{figure}[h!]
	\centering
	\includegraphics[width=9cm]{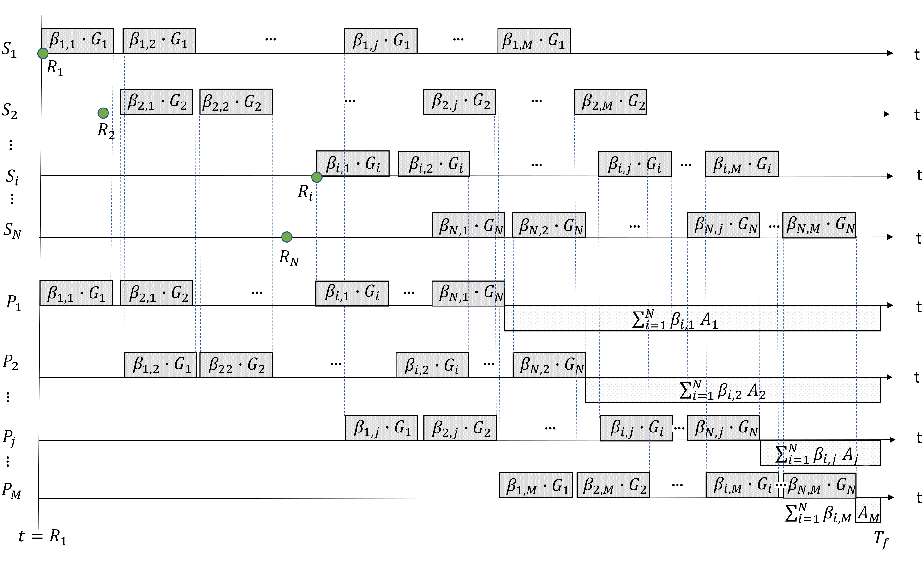}
	\caption{Timing Diagram for a Multi-source Multi-processor Network (without Front-End Processors)}
	\label{fig:ts}
\end{figure}

\subsubsection{Problem Formulation}
 
From the timing diagram it can found that gaps may appear in the communication phase. In this research, two new parameters are used to mark the time stamps of the starting and ending time of sources distributing each load fraction:\par

\begin{center}
	$TS_{i,j}$ \hspace{0.4cm} The time that source $S_{i}$ starts distributing load fraction $\beta_{i,j}$ to processor $P_{j}$.\\
	$TF_{i,j}$  \hspace{0.4cm} The time that source $S_{i}$ ends distributing load fraction $\beta_{i,j}$ to processor $P_{j}$.
\end{center}	

\textbf{A. Constraints Introduced by the Amount of Load for Each Load Fraction}\par 
	Based on the definition of $TS_{i,j}$ and $TF_{i,j}$, the following equation is used to measure the length of each load transmission between sources and processors:\par 
	\begin{eqnarray}
	\begin{aligned}
	TF_{i,j} - TS_{i,j}=\beta_{i,j}G_{i} , \\
	\hspace{0.2cm} i=1, 2, 3, ..., N, \hspace{0.2cm} j=1, 2, 3,  ..., M
	\end{aligned}
	\end{eqnarray}

 \textbf{B. Constraints Introduced by $TS_{i,j}$ and $TF_{i,j}$ on Processors}\par
	\begin{figure}[h!]
		\centering
		\includegraphics[width=8cm]{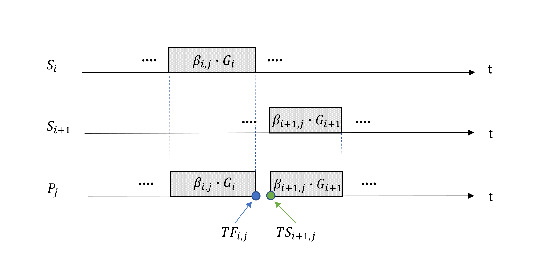}
		\caption{The Relationship Between Two Adjacent Loads Transited to Processor $P_j$}
		\label{fig:ts}
	\end{figure}
	Figure 8 shows the relationship between two adjacent load transmissions on processor $P_j$. It is clearly assumed by the sequential communication that $S_{i+1}$ has to wait until $S_i$ finishes distributing load to $P_j$:\par
	\begin{eqnarray}
	\begin{aligned}
	TF_{i,j} \leq TS_{i+1,j} , \\
	\hspace{0.4cm} i=1, 2, 3, ..., N-1, \hspace{0.4cm} j=1, 2, 3,..., M
	\end{aligned}
	\end{eqnarray}

	\textbf{C. Constraints Introduced by $TS_{i,j}$ and $TF_{i,j}$ on Sources}\par
	\begin{figure}[h!]
		\centering
		\includegraphics[width=8cm]{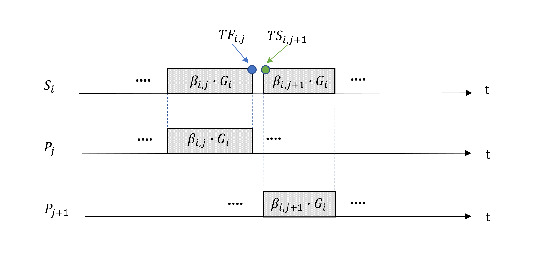}
		\caption{The Relationship Between Two Adjacent Loads Transited by Source $S_i$}
		\label{fig:ts}
	\end{figure}
	Figure 9 shows the Relationship Between Two Adjacent Loads Transited by Source $S_i$. Similar with the discussion for the last constraint, $P_{j+1}$ has to wait until $P_j$ finishes receiving load from $S_i$:\par
	\begin{eqnarray}
	\begin{aligned}
	TF_{i,j} \leq TS_{i,j+1},\\
	 \hspace{0.4cm} i=1, 2, 3,..., N, \hspace{0.4cm} j=1, 2, 3, ..., M-1
	\end{aligned}
	\end{eqnarray}

	\textbf{D. Constraints Introduced by Release Time}\par
	Firstly, the release time $R_1$ for the first source $S_1$ equals the starting time of distributing load fraction $\beta_{1,1}$:\par 
	\begin{eqnarray}
	TS_{1,1}=R_1
	\end{eqnarray}
	Secondly, the start time of the first load fraction transmission by each source should be equal or later than the release time of that source:\par 
	\begin{eqnarray}
	TS_{i,1}\geq R_i, \hspace{0.4cm} i=2, 3, 4,..., N
	\end{eqnarray}
	To make the full use of each source, it should be keep on distributing load before the next source first become available at its release time:\par 
	\begin{eqnarray}
	TF_{i-1,1}\geq R_{i}, \hspace{0.4cm} i=2, 3, 4,..., N
	\end{eqnarray}

	\textbf{E. Constraints Introduced by Finish Time}\par
	Each processing node starts processing right after it finishes receiving all the data for the sources. So the finish time of $P_j$ equals the summation of the finish time of transmitting last load fraction, which is $TF_{N,j}$, plus the computing time, $\sum_{i=1}^{N}\beta_{i,j}A_j$. Since this is an optimization problem, the finish time can be written as inequalities:\par 
	\begin{eqnarray}
	T_f\geq  TF_{N,j} + \sum_{i=1}^{N}\beta_{i,j}A_j ,\hspace{0.4cm} j=1, 2, 3,..., M
	\end{eqnarray}

	\textbf{F. Constraints Introduced by Normalization}\par
	As with the last case, the parameter total job $J$ is used to normalize all the load fractions $\beta_{i,j}$:\par 
	\begin{eqnarray}
	J=\sum_{i=1}^{N}\sum_{j=1}^{M}\beta_{i,j} 
	\end{eqnarray}

Here is the summary of the optimization problem:\par
Given the number of sources (N), number of processors (M), each source sends load fractions to all the processors in a sequence, all the sources work in parallel, find the load fractions assigned to each processor from each source such that the total system finish time is minimized.
Minimize $T_f$ such that\par 
\begin{center}

	$TF_{i,j} - TS_{i,j}=\beta_{i,j}G_{i} ,\hspace{0.2cm} i=1, 2,3, ..., N, \hspace{0.2cm} j=1, 2, 3,..., M$\\
	$TF_{i,j} \leq TS_{i+1,j} , \hspace{0.4cm} i=1, 2, 3,..., N-1, \hspace{0.4cm} j=1, 2,3, ..., M$\\
	$TF_{i,j} \leq TS_{i,j+1}, \hspace{0.4cm} i=1, 2,3, ..., N, \hspace{0.4cm} j=1, 2, 3,..., M-1$\\
	$TS_{1,1}=R_1$\\
	$TS_{i,1}\geq R_i, \hspace{0.4cm} i=2, 3,4, ..., N$\\
	$TF_{i-1,1}\geq R_{i}, \hspace{0.4cm} i=2, 3,4, ..., N$\\
	$J=\sum_{i=1}^{N}\sum_{j=1}^{M}\beta_{i,j}$\\
	$T_f\geq TF_{N,j}+\sum_{k= 1}^{N}\beta_{k,j}A_{j}\hspace{0.4cm}j=1, 2, 3,..., M$

\end{center}
Similar to the last case that already has been studied, the variables are: the system finish time $T_f$, the load fractions $\beta_{i,j},\hspace{0.2cm} i=1, 2, ..., N$, and $j=1,2, ..., M$, and the starting time and finish time for the transaction of each fraction of load $TF_{i,j}$, and $TS_{i,j}$. So it is also a problem which can be solved by linear programming techniques.\par

\section{Simulation and Numerical Tests}
This section presents multiple simulation tests to prove the improvement of multi-source multi-processor distribution system compared with regular single-source system. A numerical test will firstly be presented to show a simple case. Later, more simulations will be tested to show how the system finish time will change as the number of sources/processors increases. 
\subsection{Numerical Test}
In this numerical test, two distribution systems are created (one with front-end processor built with all processors and one without). The parameters used are listed in Table 1 and Table 2.\par  

\begin{table}[h]
	\caption{Parameters Used for the Numerical Test of Multi-source Multi-processor System (with Front-end Processors)}
	\begin{center}
		\begin{tabular}{||c c c c||} 
			\hline
			$(G_1, G_2)$ & 	$(R_1, R_2)$  &   $(A_1 ,A_2, ..., A_{5})$ & $J$  \\ 
			\hline
			(0.2, 0.4) &  (10, 50) &  (2, 3, ..., 6) & 100 \\ 
			\hline
		\end{tabular}
	\end{center}
\end{table}

	\begin{figure}[h!]
		\centering
		\includegraphics[width=9cm]{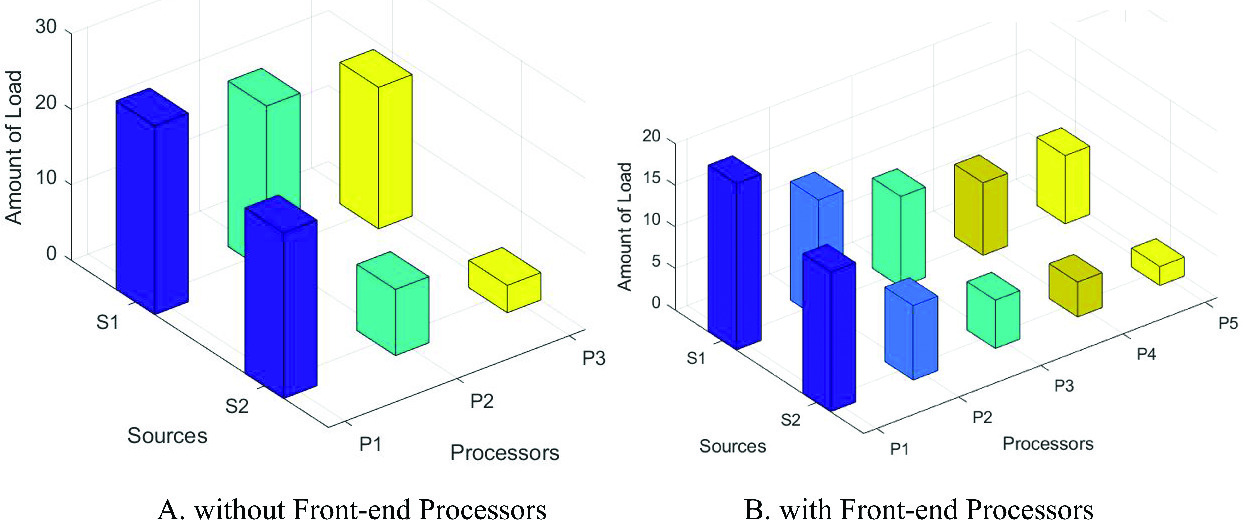}
		\caption{Amount of Load Processed by Each Processor, Received From Both Sources (with Front-end Processors)}
		\label{fig:ts}
\end{figure}
Figure 10 is the summary of load fractions that each source sends to each processor. For Figure 11, the load that both first and second source sent to each processor are added and the amount of load that each processor computes is plotted. It is very clear that the processors with faster computing speeds do more processing work than the slower ones. By using the faster processors more than the slower ones, the system can minimize the finish time.\par 

\begin{figure}[h!]
	\centering
	\includegraphics[width=8cm]{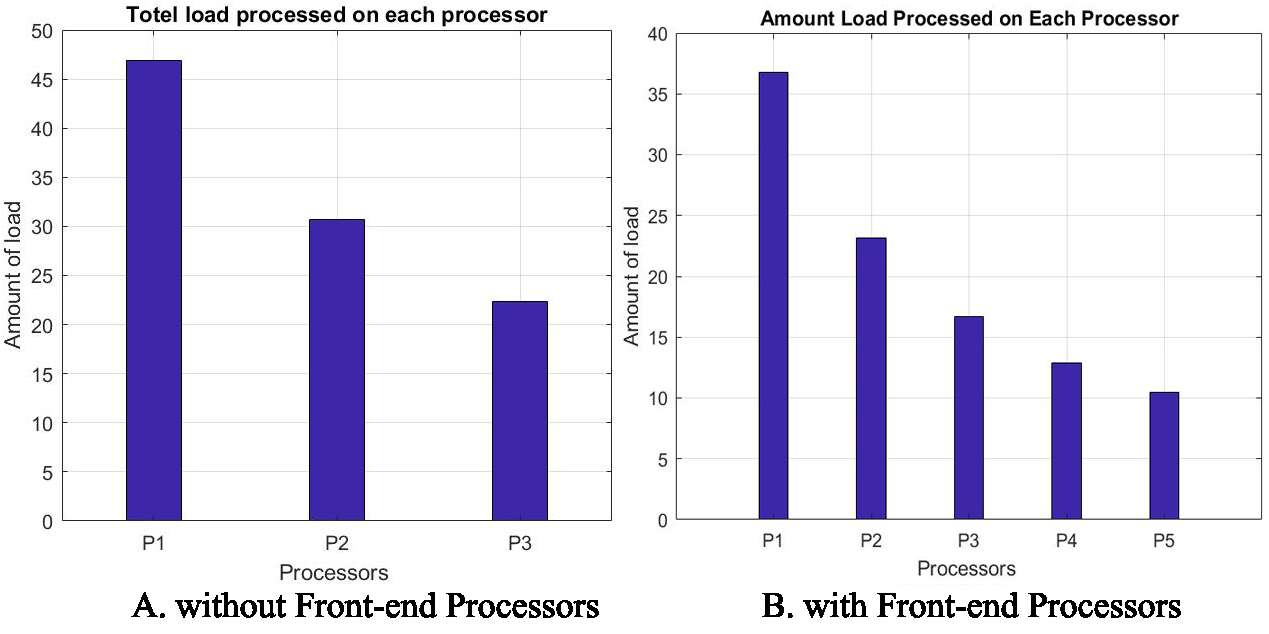}
	\caption{Amount of Load Assigned from Each Source to Each Processor (without Front-end Processors)}
	\label{fig:ts}
\end{figure}

\begin{table}[h]
	\caption{Parameters Used for the Numerical Test of Multi-source Multi-processor System (without Front-end Processors)}
	\begin{center}
		\begin{tabular}{||c c c c||} 
			\hline
			$(G_1, G_2)$ & 	$(R_1, R_2)$  &   $(A_1 ,A_2, A_3)$ & $J$  \\
			\hline
			(0.2, 0.2) &  (0, 5) &  (2, 3, 4) & 100 \\ 
			\hline
		\end{tabular}
	\end{center}
\end{table}

\subsection{Finish Time Versus Increasing Number of Sources and Processors}
	
	\begin{table}[h]
		\caption{Parameters Used for Testing the System Minimal Finish Time}
		\begin{center}
			\begin{tabular}{||c c c c||} 
				\hline
				$(G_1, G_2, G_3)$ & 	$(R_1, R_2, R_3)$  &   $(A_1 ,A_2, A_3, ..., A_{20})$ & $J$  \\ 
				\hline\hline
				(0.5, 0.6, 0.7) &  (2, 3 ,4) &  (1.1, 1.2, 1.3 ,... ,3) & 100 \\
				\hline
			\end{tabular}
		\end{center}
	\end{table}
%
	\begin{figure}[h!]
		\centering
		\includegraphics[width=8cm]{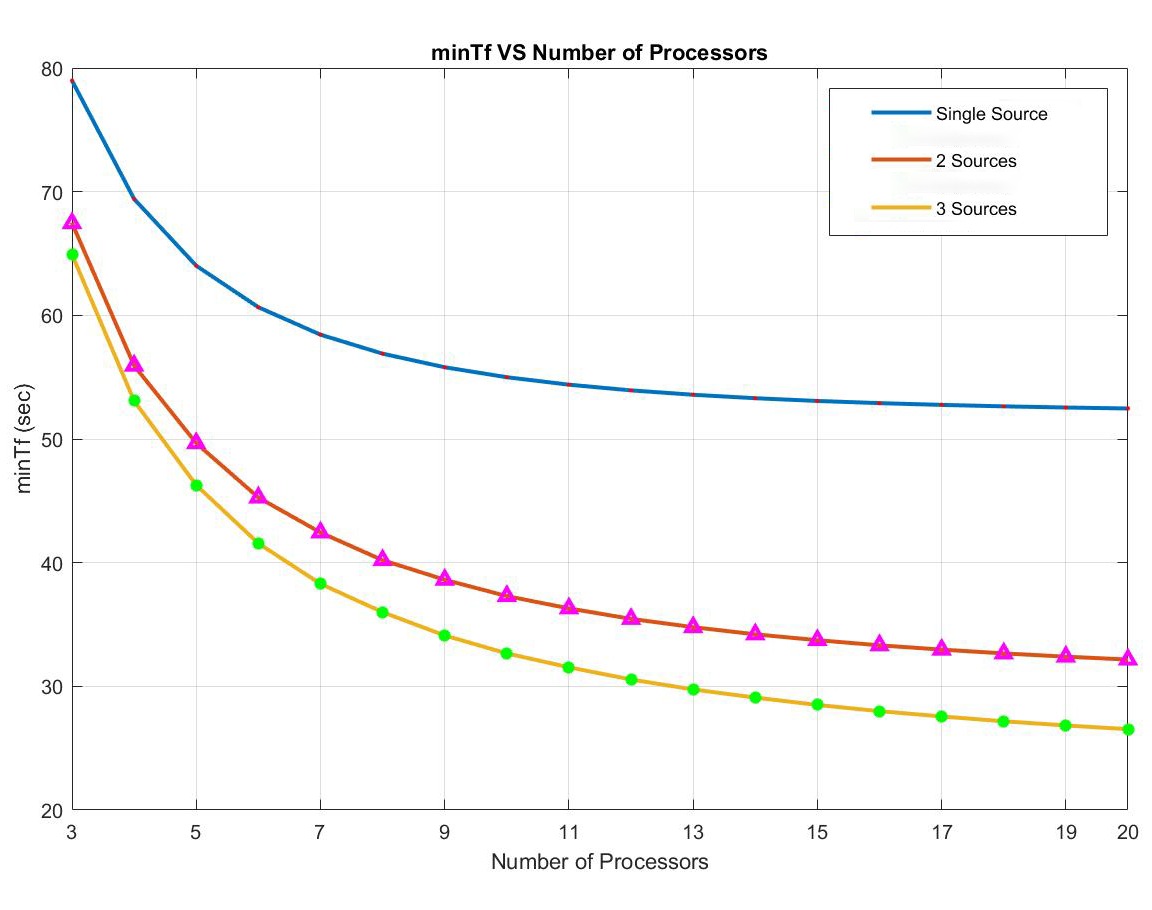}
		\caption{Minimal Finish Time vs. Number of Sources and Processors (without Front-end Processors)}
		\label{fig:ts}
	\end{figure}

	Figure 12 shows 3 cases in which the system has a single source, two sources or three sources in the system. Here all the processors are not equipped with front-end processors. The x-axis is the increasing number of processors working for the distribution system. The y-axis is the system minimal finish time in seconds. All the parameters used are the same as Table 3.\par 
As the figure shows, while adding more sources in the system, the overall finish time can be reduced since the added sources could help distributing load to the processors faster. Also, by increasing the number of processors used in  the system, finish time is also reduced. This is because more processing resources are introduced to the system and the whole system can compute the data faster. In increasing the number of processors, the influence of adding them is becoming smaller, since the new processors have slower computing speed. So compared with faster processors which were added on earlier, they can improve the system in a less significant way. The parameters used in this simulation are listed in Table 3. The simulation result for the system with front-end processors is similar to Figure 12.\par 

\subsection{Finish Time Versus Different Job Sizes}
\begin{figure}[h!]
	\centering
	\includegraphics[width=8cm]{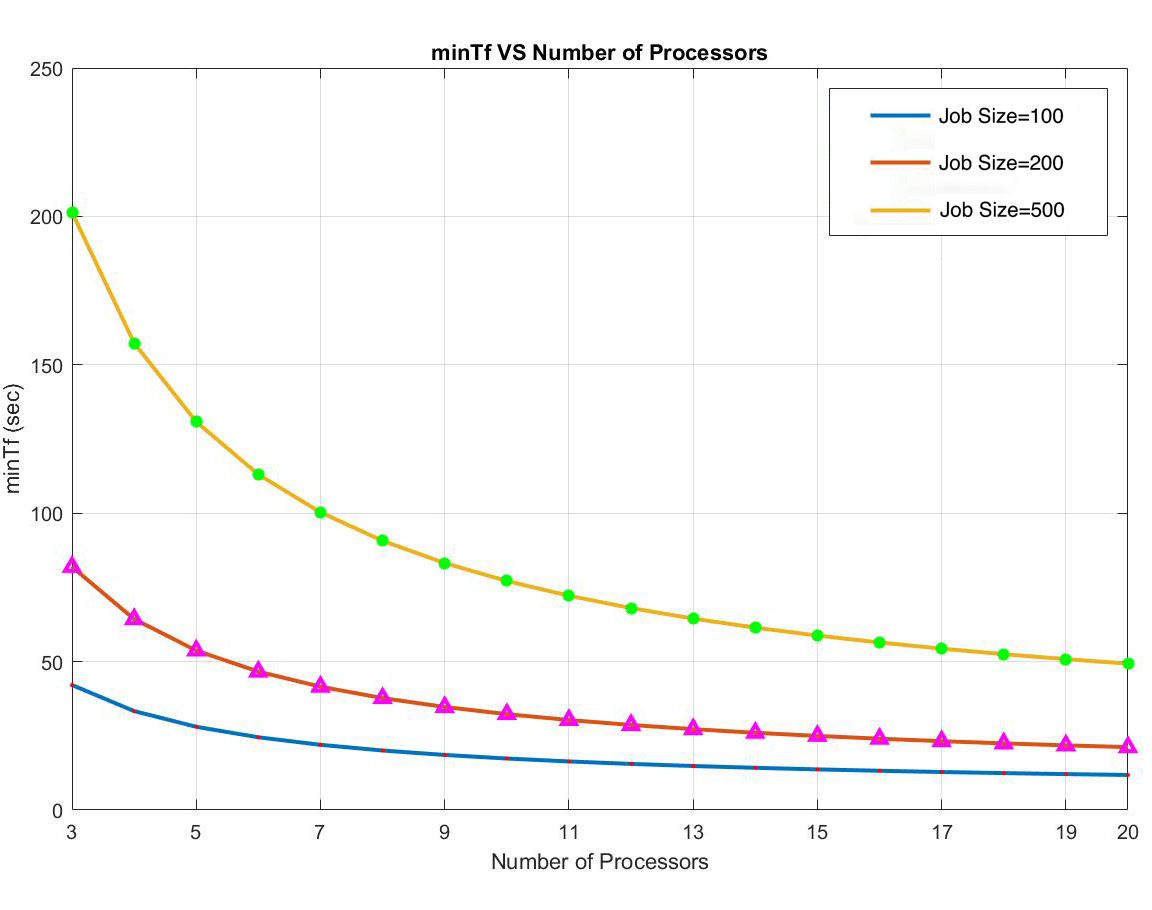}
	\caption{Minimal Finish Time vs. Number of Processors and Different Job Sizes (with Front-end Processors)}
	\label{fig:ts}
\end{figure}

Figure 13 demonstrates how the minimal finish time changes while the total job size varies. The distribution system with front-end processors was used in this simulation. There were three sources and up to 20 processors to do the simulation. The parameters used are the same as Table 3, except using three different job sizes. It is natural that the larger the job size is, the longer time the system needs to compute it. Based on the Figure 13, it can be found that the multi-source multi-processor system can have much more significant improvement while the job size is larger. For the case that the job size equals to 500, it saves about 50 percent of finish time by increasing the number of processors from three to seven. This gives us an evidence that the multi-source multi-processor job distribution system can significantly improve the performance of any large data center, sensor network, cloud network, etc.\par

\section{Speedup and System Performance Analysis}
In recent decades, Amdahl's Law is widely used as a formula to give the theoretical speedup of the execution of a task with fixed workload. Performance levels can be found by comparing different systems using Amdahl's Law with the same workload. 
\subsection{Introduction of Amdahl's Law}
Amdahl's Law was firstly created by G.H Amdahl in 1967 [20] for discussing if it was practical and efficient to use a multiplicity of processors rather than a single processor to achieve better performance. In his work [21], a performance metric called "speedup" was used to predict the theoretical speedup of execution time when using multiple processors. It is the ratio of the solution time for one processor, T(1), to the solution time for multiple processors, T(p):\par
\begin{eqnarray}
\begin{aligned}
S = \frac{T(1)}{T(p)}
\end{aligned}
\end{eqnarray}
Since the main goal for this paper is to study the improvement of using multiple sources in the load distribution system compared with the traditional single-source systems, a new equation is used to show the speedup of using p sources and n processors over q source and n processors.
\begin{eqnarray}
\begin{aligned}
S = \frac{T(1 \; source, n \;  processors)}{T(p \;  sources, n \;  processors)}
\end{aligned}
\end{eqnarray}
\subsection{Speedup Analysis and Simulations}
In this problem, both the number of source nodes and the number of processing nodes can be increased. To adapt Amdahl's Law, either the number of sources or the number of processors is fixed to compare the system optimal finish time, which can be referred to the solution time in Amdahl's Law, with the finish time using less nodes.\par
The simulation results are plotted in Figure 14 with the data in Table 4. In the simulation, the distribution system without front-end processors was used. In order to highlight the improvement of increasing the number of processors and sources, homogeneous nodes are being used during this simulation process.\par
\begin{table}[h]
	\caption{Parameters Used for the Speedup and Performance Analysis}
	\begin{center}
		\begin{tabular}{||c c c c||} 
			\hline 
			$(G_1 ,G_2, ..., G_{10})$ & 	$(R_1 ,R_2, ..., R_{10})$  &   $(A_1 ,A_2, A_3, ..., A_{18})$ & $J$   \\
			\hline\hline
			(0.5, 0.5, ..., 0.5) &  (0, 0, ..., 0) &  (2, 2, ..., 2) & 100 \\ 
			\hline
		\end{tabular}
	\end{center}
\end{table}
Figure 14 is the system finish time of the systems using 1, 2, 3, 5 and 10 sources and 1 to 18 processors. The x-axis indicates the increasing number of processors and the y-axis indicates the minimal system finish time which solved with the method discussed in previous section.\par
Figure 15 is the speedup of the system using multiple sources and processors compared with single source and the corresponding number of processors. This plot was drawn with Equation 16 and the simulation values in Figure 14. The x-axis indicates the increasing number of processors and the y-axis indicates the speedup of the system with the corresponding number of sources and processors.\par
Figure 15 shows that by adding more sources to the system, the speedup becomes larger. For example, the speedup for the system of 2 sources and 12 processors is around 1.59, comparing with the speedup using 3 sources/ 5 sources/ 10 sources (also 12 processors) to be 1.90/ 2.21/ 2.49. In this example, the speedup value of using 3 sources has an improvement of 19\% compared with the case using 2 sources.  The speedup value of using 10 sources has an improvement of 57\%.\par
The relative low values of speedup observed here are due to inefficiencies of the sequential distribution protocol [5].\par 
Meanwhile, by observing each fitted line of the speedup values, one can easily see that the speedup value for using the same number of source and increasing number of processors is also gradually getting larger. \par
These observations further prove that the multi-source multi-processor system provides improvement for the load distribution system by reducing the system minimal finish time and boosting the system speedup level.
\begin{figure}[h!]
	\centering
	\includegraphics[width=9cm]{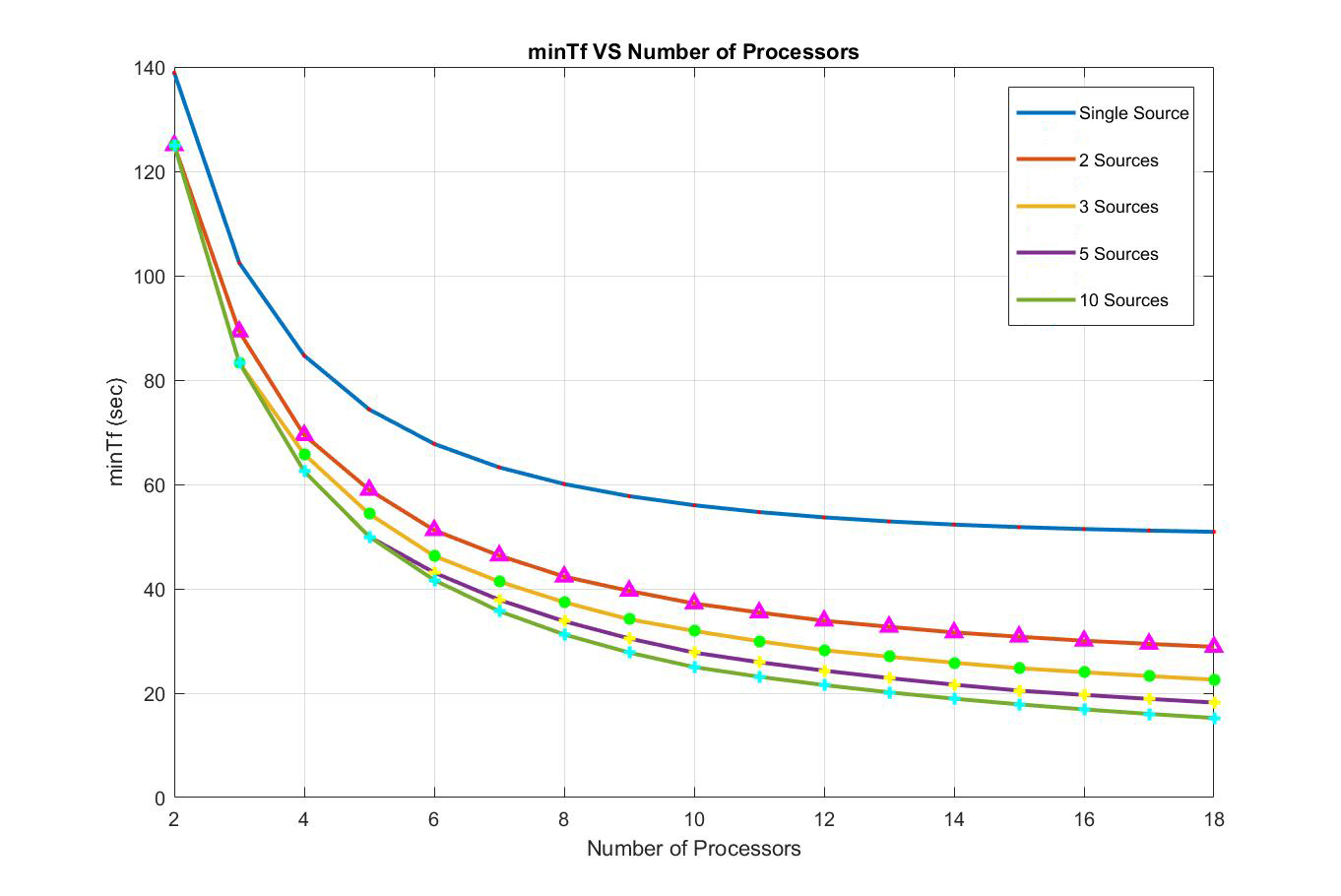}
	\caption{Minimal Finish Time for the System without Front-end Processors}
	\label{fig:ts}
\end{figure}
\begin{figure}[h!]
	\centering
	\includegraphics[width=9cm]{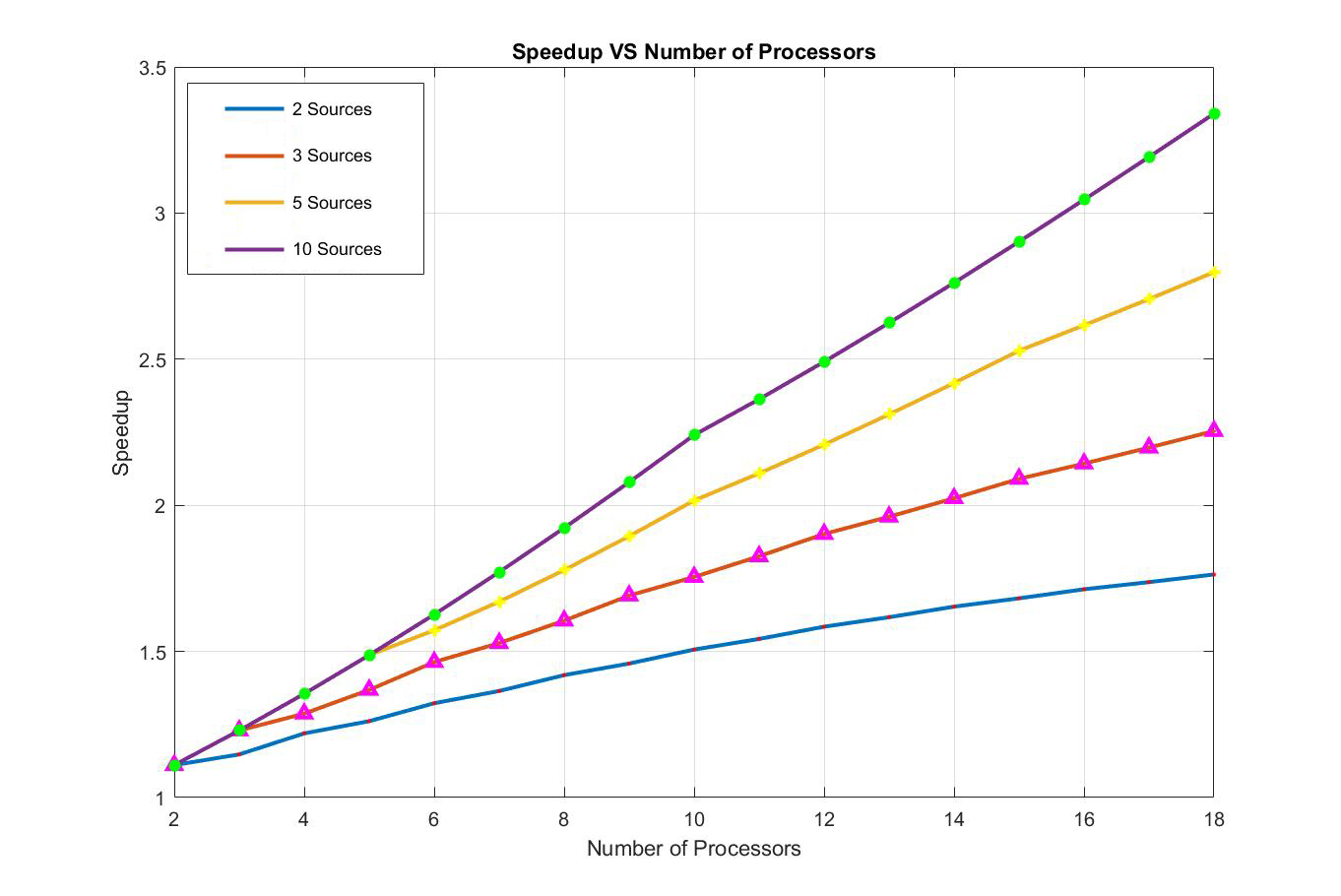}
	\caption{Speedup for the System without Front-end Processors}
	\label{fig:ts}
\end{figure}
\section[Trade-off Analysis]{Trade-off Analysis  for Minimal Finish Time and Monetary Cost}
In this section, since the computing of jobs requires a great amount of computing power, a concept called monetary cost is introduced to measure the monetary cost for hiring the  processors' computing power. Monetary cost was previously studied in [23] [24]. Here a trade-off analysis is presented with several suggestive plans given to users who have budgets on monetary cost, or have to finish processing the total data within a certain finish time, or have the budget for both money and time. In this section all the simulations are done with the network with front-end processors equipped with the processing nodes.
\subsection{Definition of Monetary Cost}
The term monetary cost is the cost for using sources or processors for processing the load. This paper mainly focus on the monetary cost for the processors. The monetary cost for processors $P_j$ is defined as $C_{j}$. The unit for them is cost/unit time. So the total cost for $P_j$ to process load fraction $\beta_{i,j}$ is $\beta_{i,j}A_jC_{j}$.
The total monetary cost for the entire system to finish processing job $J$ is:\par 
\begin{eqnarray}
Cost_{total}=\sum_{i=1}^{N}\sum_{j=1}^{M}\beta_{i,j}A_jC_{j}
\end{eqnarray}
In this paper, there is an assumption that the faster processors have more expensive monetary cost, which is: $C_1>C_2>...>C_M$. 
\subsection{Trade-off Analysis with a Cost Budget}
This section is going to discuss how many processors should be used given a cost budget $Budget_{cost}$. Since both the number of sources and the number of processors influences the results, and this paper is mainly discussing the computing cost given by the processors. In the case, the number of sources is fixed to be two. The parameters used in this section are listed in Table 6.\par 
\begin{table}[h]
	\caption{Parameters Used for the Trade-off Analysis}
	\begin{center}
		\begin{tabular}{||c c c c c||} 
			\hline
			$(G_1, G_2)$ & 	$(R_1, R_2)$  &   $(A_1 ,A_2, ..., A_{20})$ & $(C_1, C_2, ..., C_{20})$ & $J$  \\
			\hline\hline
			(0.5, 0.6) &  (2, 3) &  (1.1, 1.2,  ..., 3) & (29, 28,  ..., 10) & 100 \\ 
			\hline
		\end{tabular}
	\end{center}
\end{table}

	\textbf{STEP 1. Plot the Number of Processors VS. Total Cost}\par  
	\begin{figure}[h!]
		\centering
		\includegraphics[width=9cm]{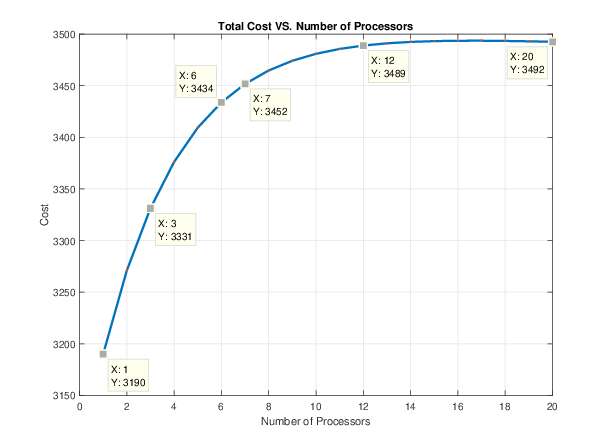} 
		\caption{System Total Monetary Cost of Computing VS. Number of Processors}
		\label{fig:ts}
	\end{figure}
	First, the relationship between the number of processors and total cost is plotted as Figure 16. The x-axis is the number of processors used in the distribution system, and y-axis is the total cost for computing, where the units are dollars. It is natural that the total computing cost is growing as the number of processors increases. However the growth rate is becoming smaller. This is because although more processors are used in the system, while solving the optimal problem of finish time, the slower/cheaper processors are assigned with much less amount of load to process. This makes them have less influence on the total cost. \par 
	As an example, given that the budget for the system monetary cost is $Budget_{cost}=$3450 dollars. By looking into the list of total cost, the two closest solutions can be found:\par
	Using 6 processors: the total computing cost is about 3433.77 dollars;\par 
	Using 7 processors: the total computing cost is about 3451.67 dollars.\par 
	In this case, all the solutions that using less than or equal to 6 processors is within the budget of 3450 dollars.\par
	
	\textbf{STEP 2. Plot the Number of Processors VS. Gradient of Finish Time and the Gradient of  $T_f$}\par 
	\begin{figure}[h!]
		\centering
		\includegraphics[width=9cm]{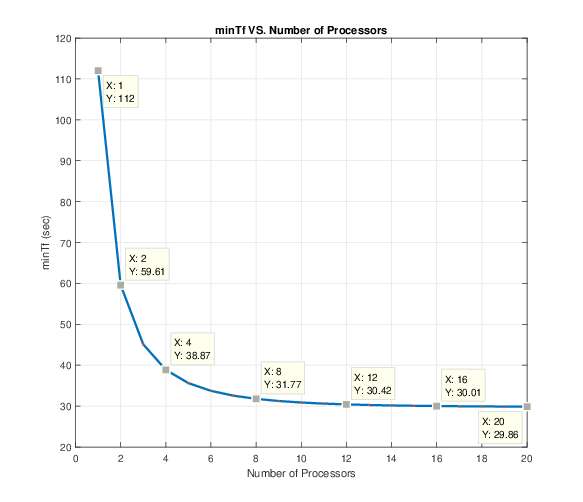}
		\label{fig:ts}
		\caption{System Minimal Finish Time VS. Number of Processors}  
	\end{figure}
	Second, Figure 17 is plotted to show the relationship between the number of processors and system minimal finish time. 
	\begin{figure}[h!]
		\centering
		\includegraphics[width=9cm]{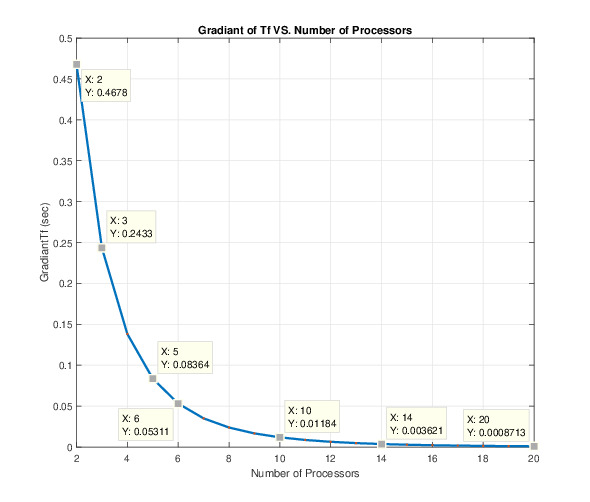}
		\caption{Gradient of System Minimal Finish Time VS. Number of Processors}  
		\label{fig:ts}
	\end{figure}
	Figure 18 shows the gradient of finish time. The definition of the gradient of $T_f$ is:\par 
	\begin{eqnarray}
	Gradiant_{T_{f,m}}=\frac{T_{f,m\hspace{0.1cm}processors}-T_{f,m-1\hspace{0.1cm}processors}}{T_{f,m-1\hspace{0.1cm} processors}}
	\end{eqnarray}
	The values of the gradient of finish time demonstrates by what percentage can the new solution make the whole system finish the job faster. In this test result, $Gradient_{T_{f,5}}$ is about 8.4\%, and $Gradient_{T_{f,6}}$ is about 5.3\%. \par 
	
	\textbf{STEP 3. The Trade-off Plan}\par 
	Now let us discuss a trade-off plan. It is clear that when the number of processors increases, the finish time decreases but the monetary cost increases. So there must be a trade-off between finish time and monetary cost. Suppose when adding one more processor to the system, the finish time is reduced by less than 6\%, then the user may prefer using less processors to reduce the monetary cost rather than using one more processor to reduce finish time, which is already reaching a very low value. In this way, a good suggestion can be given to the user about how many processors should be used in the system to be within the budget of cost. In this example, the user should use 5 processors.

\subsection{Trade-off Analysis with a Time Budget}
This section is going to discuss how many processors should be used given the maximal of time that the total job needs to be finished distributing and processing, which is called $Budget_{time}$. The simulation results in the last section is used here.\par 
First, the user increases the number of processors from 1 to $m$, where $T_{f,m\hspace{0.1cm}processors}\leq Budget_{time}$, and $T_{f,m-1\hspace{0.1cm}processors}\geq Budget_{time}$. Since the finish time decreases as the number of processors increases, and $Budget_{time}$ is the maximum finish time that the user required, all the solutions that have more than $m$ processors could meet the requirement. For example, while $Budget_{time}$=32 seconds,  all the solutions with equal to or more than 10 processors meets the requirement.\par 
Meanwhile, since the total computation monetary cost increases as the number of processors increases, with the purpose of saving money, the user should use as few processors as possible, which is $m$ processors. In the example, the user should pick 10 processors.

\subsection{Trade-off Analysis with Both a Time Budget and a Cost Budget}
In this section, both the time budget and the cost budget are considered. By combining the two graphs of number of processors versus finish time and total cost, the solution area which meet both of the requirements is highlighted.\par
	 \textbf{CASE 1. The Two Solution Areas Overlapped}\par 
	\begin{figure}[h!]
		\centering
		\includegraphics[width=9cm]{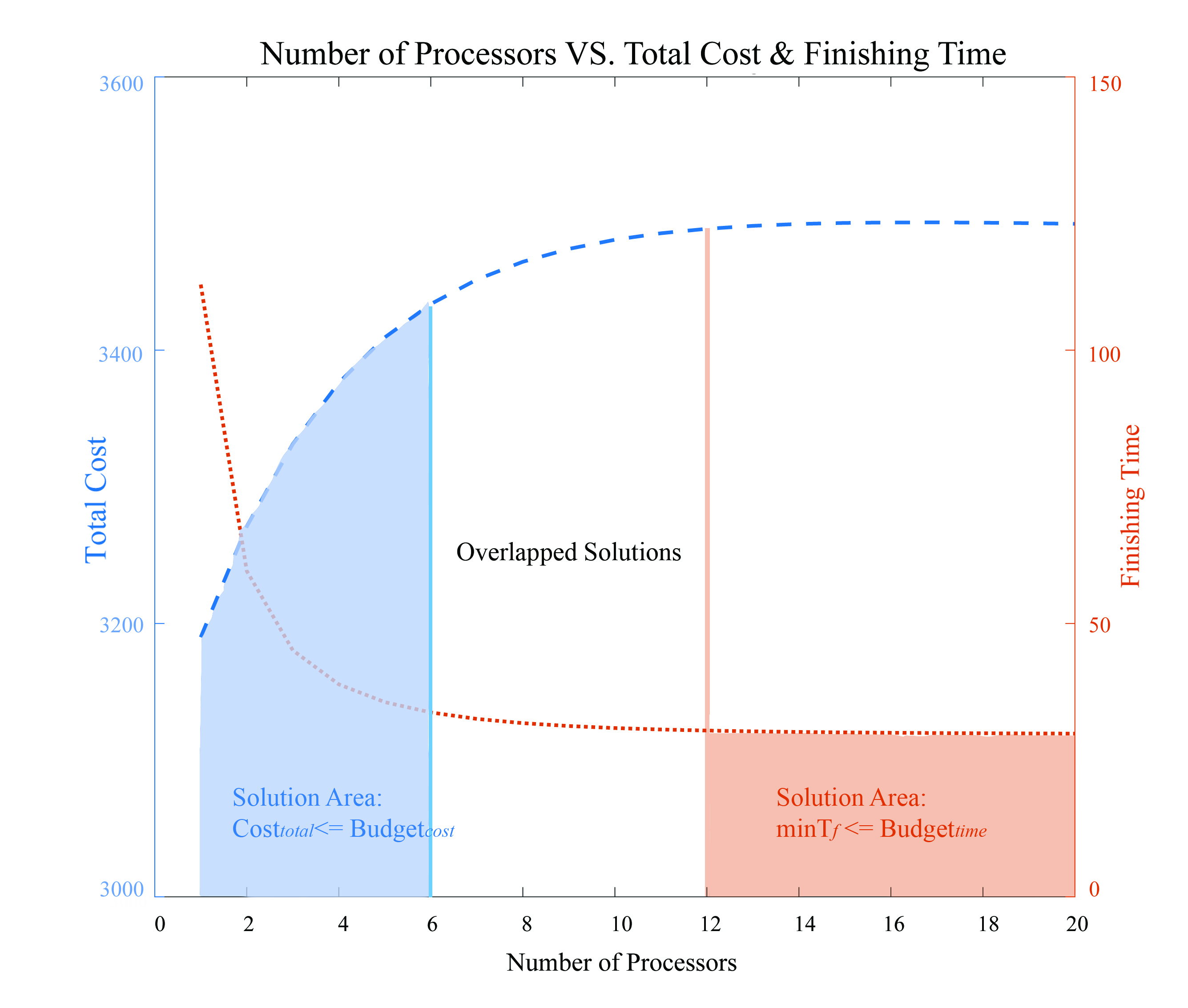}
		\caption{Two Solution Areas with Both a Time Budget and a Cost Budget: Case 1, Solution Areas are Overlapped}
		\label{fig:ts}
	\end{figure}
	In Figure 19, the x-axis is the number of processors involved in the test, the left y-axis is the total cost for the processors, and the right y-axis is the overall system minimal finish time. As Figure 19 shows, the solution area of $Cost_{total}\leq Budget_{cost}$ is highlighted in blue, and the solution area of $minT_{f}\leq Budget_{time}$ is highlighted in orange. They have a overlapped solution area, where the number of processors $m$ can varies from 6 to 12. All the systems from 6 to 12 processors satisfy both the cost budget and the time budget.  \par
	\textbf{CASE 2. There is No Overlap Between Two Solution Areas}\par 
	\begin{figure}[h!]
		\centering
		\includegraphics[width=9cm]{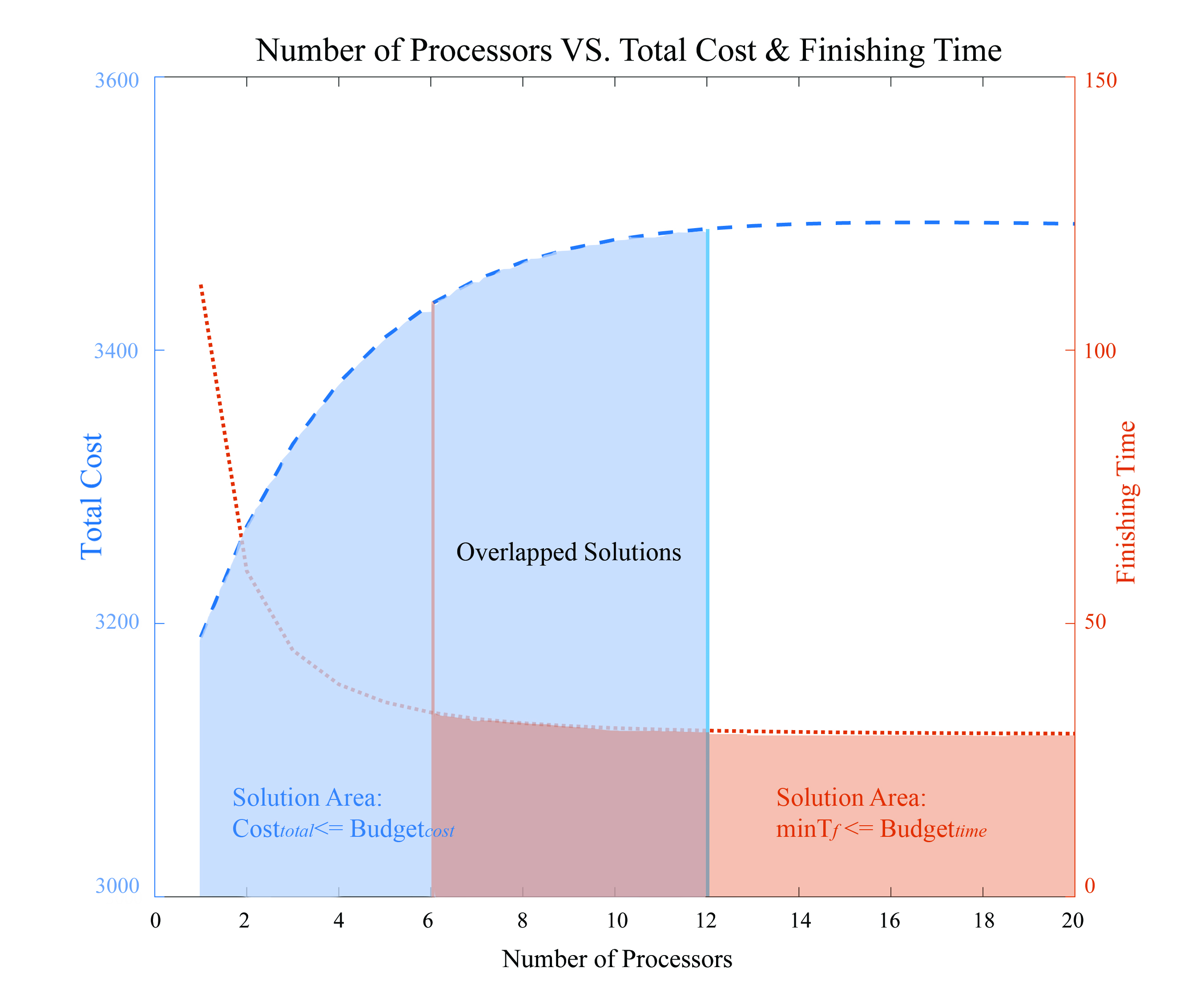}
		\caption{Two Solution Areas with Both a Time Budget and a Cost Budget: Case 2, Solution Areas have No Overlap}
		\label{fig:ts}
	\end{figure}
	Same as in the last figure, for Figure 20, the x-axis is the number of processors involved in the test, the left y-axis is the total cost for the processors, and the right y-axis is the overall system minimal finish time. In Figure 20, the two solution areas  there $Cost_{total}\leq Budget_{cost}$ and $minT_{f}\leq Budget_{time}$ are highlighted. Since there is no overlap between these two areas,  there is no solution to satisfy both the cost budget and the time budget. The user has to either increase the amount of money to spend on processing the whole job, or wait a longer time for the system to finish processing. \par

\section{Conclusion}
This paper studies the load distribution and finish time optimization problem for a multi-source, multi-processor network based on the two-level tree network topology. The study was divided into two scenarios: the processing nodes are equipped with or without front-end processors. Numerical tests and simulations results showed that the multi-source system has great improvement compared with single-source system, since the overall system minimal finish time is reduced significantly. Meanwhile, by increasing either the number of sources or processors, the finish time can be further reduced. Then, a monetary cost model is proposed to calculate the computing power used for the system. Finally, since monetary cost and minimal finish time has a trade-off relationship, three trade-off plans are demonstrated: 1. the user has a cost budget; 2. the user has a time budget; 3. the user has both a cost budget and a time budget.
\section{Future Work}
In this paper, it is assumed that if the source or processor has to communicate with multiple nodes, it uses sequential communication. This means that the  source or processor could only communicate with one node at a time. However, with the rapid growth of the technology, it is very common to use the sources and processors which can do simultaneous communication with a bandwidth limitation. In the future work, the bandwidth parameters should to be modified to see how much faster the system can be improved.\par
On the other hand, a more complicated but realistic scenario may have multiple jobs arrive at the processing nodes during the processing phase, which makes the processing speed become time-varying. Also, the sources' communication speed can also be time-varying due to the injection of some job distributing tasks. It will be a very valuable study to combine the current study with this scenario.\par 
Another interesting topic is the combination of Divisible Load Theory and Amdahl's Law. Amdahl's law is a formula used to find the maximum improvement possible by improving a particular part of a system. In parallel computing, Amdahl's law is mainly used to predict the theoretical maximum speedup for program processing using multiple processors. Since it is a very useful tool for predicting speedup, with the help of it, new methods can be discovered to improve parallel systems while adapting our current study to more complicated network topologies.\par

\ifCLASSOPTIONcaptionsoff
  \newpage
\fi

\begin{IEEEbiography}[{\includegraphics[width=1in,height=1.25in,clip,keepaspectratio]{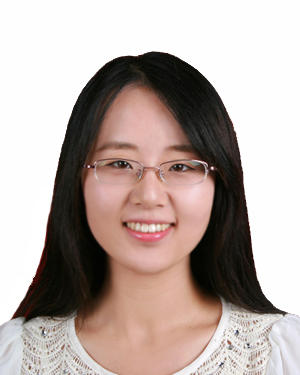}}]{Yang Cao}
	received the BE degree in Electrical Engineering and Automation from Northwestern Polytechnical University, Xi'an, China, in June 2012. She also received MS degree in Electrical Engineering from Stony Brook University, Stony Brook, New York, in December 2013. Currently she is working toward the PhD degree in Electrical Engineering at Stony Brook University. Her research interests include task scheduling and resource allocation in distributed systems, cloud networks,  data centers, etc.
\end{IEEEbiography}
\begin{IEEEbiography}[{\includegraphics[width=1in,height=1.25in,clip,keepaspectratio]{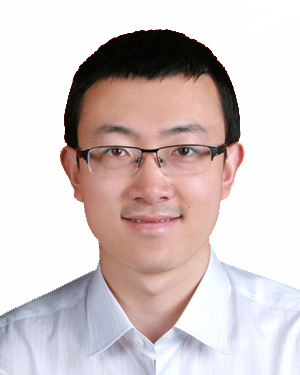}}]{Fei Wu}
	received the BE degree in information and telecommunication engineering from Xi'an Jiaotong University, Xi'an, China, in 2012, and the MS degree in electrical engineering from Stony Brook University, Stony Brook, New York, in 2013. He is currently working toward the PhD degree in electrical engineering at Stony Brook University. His research interests include scheduling, parallel processing, computer networks and virtualization. 
\end{IEEEbiography}

\begin{IEEEbiography}[{\includegraphics[width=1in,height=1.25in,clip,keepaspectratio]{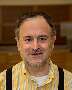}}]{Thomas G. Robertazzi}
	received the BEE degree from Cooper Union, New York, in 1977 and the PhD degree from Princeton University, Princeton, New Jersey, in 1981. He is presently a professor in the Department of Electrical and Computer Engineering, Stony Brook University, Stony Brook, New York. He has published extensively in the areas of parallel processing scheduling, telecommunications and performance evaluation. He has also authored, co-authored or edited six books in the areas of networking, performance evaluation, scheduling and network planning. He is a fellow of the IEEE and since 2009 co-chair of the Stony Brook University Senate Research Committee.
\end{IEEEbiography}


\begin{thebibliography}{1}
\bibitem{1}
Cheng, Y.C. and T. G. Robertazzi, \emph{Distributed Computation with Communication Delays}, IEEE Transactions on Aerospace and Electronic Systems, Volume. 24, Issue. 6, Nov. 1988, pp. 700 - 712.
\bibitem{2}
Sohn, J. and T. G. Robertazzi, \emph{Optimal Load Sharing for a Divisible Job on a Bus Network}, IEEE Transactions on Aerospace and Electronic Systems, Vol. 32, Issue. 1, Jan. 1996, pp. 34 - 40. 
\bibitem{3}
Kim, H.J., Jee, G.-I. and Lee, J.G., \emph{Optimal Load Distribution for Tree Network Processors}, IEEE Transactions on Aerospace and Electronic Systems, Vol. 32, Issue. 2, April 1996, pp. 607 - 612.
\bibitem{4}
M. Drozdowski and W. Glazek, \emph{Scheduling Divisible Loads in a Three-Dimensional Mesh of Processors}, Parallel Computing, Volume. 25, Issue. 4, April 1999, pp. 381 - 404.
\bibitem{5}
B. Veeravalli, D. Ghose, V. Mani and T. G. Robertazzi, \emph{Scheduling Divisible Loads in Parallel and Distributed Systems}, IEEE Computer Society Press, 1996.
\bibitem{6}
L. Ping, B. Veeravalli and A. A. Kassim,  \emph{Design and implementation of parallel video encoding strategies using divisible load analysis}, IEEE Transactions on Circuits and Systems for Video Technology, Vol. 15, Issue: 9, Sept. 2005, pp. 1098 - 1112.
\bibitem{7}
X. Li, X. Liu and H. Kang, \emph{Sensing Workload Scheduling in Sensor Networks Using Divisible Load Theory} Global Telecommunications Conference, 2007. GLOBECOM ’07. IEEE, 2007, pp. 785 – 789.
\bibitem{8}
X. Li, H. Kang and J. Cao, \emph{Coordinated Workload Scheduling in Hierarchical Sensor Networks for Data Fusion Applications}, Journal of Computer Science and Technology, Volume. 23, 2008, pp. 355 - 364.
\bibitem{9}
Daniel, D., and Lovesum, S. P. J. \emph{A Novel Approach for Scheduling Service Request in Cloud with Trust Monitor}, IEEE International Conference on Signal Processing, Communication, Computing and Networking Technologies, 2011, 509 – 513.
\bibitem{10}
Wang, X., Wang, B., and Huang, J., \emph{Cloud computing and its key techniques}, Computer Science and Automation Engineering (CSAE), Volume. 3. IEEE, 2011, pp. 404 – 410.
\bibitem{11}
Yang, Y., Choi, J. Y., Choi, K. M., Gannon, P. M., and Kim, D. S. \emph{BioVLAB-Microarray: Microarray Data Analysis in Virtual Environment}, Proc. IEEE E-science, 2008, 159 – 165.
\bibitem{12}
Armbrust, M., Fox, A., Griffith, R., Joseph, A. D., Katz, R. H., Konwinski, A., Lee, G., Patterson, D. A., Rabkin, A., Stoica, I., and Zaharia, M.  \emph{Above the Clouds: A Berkeley View of Cloud Computing}, University of California, Berkeley, Tech. Rep. No. UCB/EECS-2009028, 2009.
\bibitem{13}
Mell, P., and Grance, T.  \emph{The NIST Definition of Cloud Computing}, NIST Special Publication 800-145. NIST, US Department of Commerce, 2011.
\bibitem{14}
H. M. Wong, D. Yu, B. Veeravalli and T. G. Robertazzi, \emph{Data Intensive Grid Scheduling: Multiple Sources with Capacity Constraints}, Proc. 15th Int’l Conf. Parallel and Distributed Computing and Systems, 2003.
\bibitem{15}
D. Yu and T.G. Robertazzi, \emph{Multi-Source Grid Scheduling for Divisible Loads}, Proc. 40th Annual Conference on Information Sciences and
Systems, 2006. IEEE, 2006, pp. 188 – 191.
\bibitem{16}
M. Abdullah, M. Othman, \emph{Cost-Based Multi-QoS Job Scheduling using Divisible Load Theory in Cloud Computing},  Procedia Computer Science, Volume 18, 2013, Pages 928 - 935.
\bibitem{17}
S. Suresh, H. Huang, H. J. Kim, \emph{Scheduling in Compute Cloud With Multiple Data Banks Using Divisible Load Paradigm}, IEEE Transactions On Aerospace And Electronic Systems Volume. 51, No. 2 April 2015, pp. 1288 – 1297.
\bibitem{18}
A. Shokripour, M. Othman, \emph{Survey on Divisible Load Theory and its Applications}, International Conference on Information Management and Engineering, 2009, pp 300 - 304. 
\bibitem{19}
T.G. Robertazzi, \emph{Ten Reasons to Use Divisible Load Theory}, Computer, Volume. 36, Issue. 5, May 2003 , pp.63 - 68.
\bibitem{20}
G.M. Amdahl, \emph{Validity of the Single Processor Approach to Achieving Large Scale Computing Capabilities}, Proceedings of the AFIPS Conference, 1967, pp. 483-485.
\bibitem{21}
G.M. Amdahl, \emph{Computer Architecture and Amdahl's Law}, Computer, Volume. 46, Issue. 12, Dec. 2013, pp. 38 - 46.
\bibitem{22}
Agrawal, R. and Jagadish, H.V., \emph{Partitioning Technqiues for Large-Grained Parallelism}, Proceedings of the Seventh Annual International Phoenix Conference on Computers and Communications, March 1988, pp. 31 - 38.
\bibitem{23}
Sohn, J., Robertazzi, T.G. and Luryi, S., \emph{Optimizing Computing Costs using Divisible Load Analysis}, IEEE Transactions on Parallel and Distributed Systems, Volume. 9, No. 3, March 1998, pp. 225 - 234.
\bibitem{24}
Charcranoon, S., Robertazzi, T.G. and Luryi, S., \emph{Parallel Processor Configuration Design with Processing/Transmission Costs}, IEEE Transactions on Computers. Volume. 49, No. 9, Sept. 2000, pp. 987 - 991.
\end{thebibliography}
\end{document}